\documentclass[12pt,eqsecnum]{article}
\usepackage[dvips]{graphicx}
\usepackage{amssymb}
\usepackage{amsmath}
\usepackage{epstopdf}
\usepackage{color}
\usepackage{soul}
\usepackage{ulem}
\makeatletter

\@addtoreset{equation}{section}
\makeatletter

\newcommand{\ma}[1]{\mbox{$\mathcal{#1}$}}
\newcommand{\qed}{\hbox{\rule[-2pt]{6pt}{6pt}}}
\newcommand{\D}{{\rm d}}

\setstcolor{red}

\newtheorem{Prop}{Proposition}

\newcommand{\dalm}{\kern1pt\vbox{\hrule height 0.9pt\hbox{\vrule width
0.9pt\hskip 2.5pt\vbox{\vskip 5.5pt}\hskip 3pt\vrule width 0.3pt}\hrule height
0.3pt}\kern1pt}

\def\b2hat{ {\hat b}_2 }

\def\be {\begin{equation}}
\def\ee  {\end{equation}}
\def\bea {\begin{eqnarray}}
\def\eea {\end{eqnarray}}

\begin{document}

\begin{titlepage}
\vfill
\begin{flushright}
\today
\end{flushright}

\vfill
%\vskip 1.0cm
\begin{center}
\baselineskip=16pt
{\Large\bf 
Higher-dimensional Buchdahl and Janis-Robinson-Winicour transformations in the Einstein-Maxwell system with a massless scalar field\\
}
\vskip 0.5cm
{\large {\sl }}
\vskip 10.mm
{\bf  Hideki Maeda${}^{a}$ and Cristi{\'a}n Mart\'{\i}nez$^{b}$} \\

\vskip 1cm
{
   	${}^a$ Department of Electronics and Information Engineering, Hokkai-Gakuen University, Sapporo 062-8605, Japan.\\
    ${}^b$ Centro de Estudios Cient\'{\i}ficos (CECs), Av. Arturo Prat 514, Valdivia, Chile. \\
	\texttt{h-maeda@hgu.jp, martinez@cecs.cl}

     }
\vspace{6pt}
%\today
\end{center}
\vskip 0.2in
\par
\begin{center}
{\bf Abstract}
\end{center}
\begin{quote}
We present higher-dimensional generalizations of the Buchdahl and Janis-Robinson-Winicour transformations which generate static solutions in the Einstein-Maxwell system with a massless scalar field.
While the former adds a nontrivial scalar field to a vacuum solution, the latter generates a charged solution from a neutral one with the same scalar field.
Applying these transformations to  (i) a static solution with an Einstein base manifold, (ii) a multi-center solution, and (iii)  a four-dimensional cylindrically symmetric solution, we construct several new exact solutions.
\vfill
% \hrule width 5.cm
\vskip 2.mm
\end{quote}
\end{titlepage}

%<<<<<<<<<<<<< PACS NUMBER >>>>>>>>>>>>>>>%
%\pacs{
%04.50.-h 	Higher-dimensional gravity and other theories of gravity
%04.50.Gh 	Higher-dimensional black holes, black strings, and related objects 
%04.60.-m 	Quantum gravity
%04.60.Ds 	Canonical quantization 
%04.60.Kz 	Lower dimensional models; minisuperspace models 
%} 

% CECS-PHY-13/09

%\maketitle
%\section{}
%\subsection{}

\tableofcontents

\newpage 

%======================================%
%<<<<<<<<<<<< SECTION I  >>>>>>>>>>>>>>%
%======================================%
\section{Introduction}

As stationary spacetimes are very fundamental configurations of gravity, a classification of stationary solutions is one of the most important problems in the study of exact solutions in general relativity.
In the case where spacetimes possess spherical symmetry, the general vacuum solution is given by the Schwarzschild solution without assuming staticity as a consequence of the Birkhoff's theorem.
This theorem can be extended in the presence of a Maxwell field and then the general electrically charged solution consists of the Reissner-Nordstr\"om solution and the Bertotti-Robinson solution.
In the system with a scalar field, in contrast, the Birkhoff's theorem does not hold in general because it introduces dynamical degrees of freedom of the radially oscillating gravitational wave into the system.
However, if one assumes staticity, the most general spherically symmetric solution with a massless scalar field is the so-called Janis-Newman-Winicour (JNW) solution~\cite{jnw1968}\footnote{Actually, Fisher found this solution for the first time~\cite{Fisher1948} and subsequently it has been rediscovered by different authors many times~\cite{jnw1968,bl1957,Buchdahl1959,wyman1981}.} 

The above results with a massless scalar field can be generalized in the arbitrary $n(\ge 4)$-dimensional spacetime with an Einstein base manifold.
Contrary to this, it has been recently clarified that static solutions are quite rich even in such a symmetric spacetime if there exists a nontrivial Maxwell field in addition~\cite{mm2016}.
Especially, while the asymptotically flat solution is unique when either a Maxwell field or massless scalar field is trivial, there are more than one asymptotically flat solutions if both of them are nontrivial.
In addition, such a system also admits asymptotically Bertotti-Robinson solutions.

In such a classification of static solutions, solution-generating methods have been playing a central role with brilliant success.
In the four-dimensional system with a Maxwell field and a massless scalar field, the Buchdahl transformation~\cite{Buchdahl1959} and the Janis-Robinson-Winicour (JRW) transformation~\cite{jrw1969} are known as methods to generate static solutions.
While the Buchdahl transformation generates a solution with a nontrivial scalar field from a vacuum solution, the JRW generates electrically charged solution from a neutral one keeping the same form of the scalar field.

In the case with spherical symmetry, the Buchdahl transformation generates the JNW solution from the Schwarzschild solution~\cite{jrw1969}.
This Buchdahl transformation has been generalized in higher dimensions by Tangen as a special case of the more general transformation~\cite{tangen2007}, but the procedure of his higher-dimensional version is more complicated than the original four-dimensional one.
Nonetheless, the transformation certainly generates the higher-dimensional JNW solution~\cite{JNWhigher} from the Schwarzschild-Tangherlini solution~\cite{wr2007}.
This result is reasonable because both of those solutions are the general static solutions in each system.
On the contrary, as shown by the authors~\cite{mm2016}, the general spherically symmetric and static solution is not unique with a Maxwell field in addition.

Then a natural question arises: Which charged solution in the complete classification in~\cite{mm2016} is obtained from the JNW solution by the JRW transformation?
Moreover, the counterpart of the JRW transformation in higher dimensions is still missing at present.
We will answer these questions in the present paper.
In the following section, after presenting a different and much simpler formulation of the higher-dimensional Buchdahl transformation than the one in the article~\cite{tangen2007}, we will establish the higher-dimensional JRW transformation in a similar manner. 
There we also present an alternative derivation of the JRW transformation based on the reduced action for static spacetimes in the form of a nonlinear sigma model. 
In Section~\ref{sec:application}, adopting the higher-dimensional JRW transformation to the generalized JNW solution with an Einstein base manifold characterized by its curvature $k$, we will show that it is transformed into the type-I solution for $k=1,-1$ and the type-VI${}_0$ solution for $k=0$ in our classification~\cite{mm2016}.
Also, as another demonstration, we will construct new neutral and electrically charged multi-center solutions with a ghost scalar field.  As the third application, we will use the Buchdahl for dressing the four-dimensional cylindrically symmetric Levi-Civita solution, and then the JRW transformation to obtain a new charged cylindrically symmetric solution with a massless scalar field.
Concluding remarks and future prospects are given in the final section.

Our basic notations follow~\cite{wald}.
Throughout in this article, the Minkowski metric has the signature $(-,+,\cdots,+)$.
We adopt the units such that $c=1$ and the $n$-dimensional gravitational constant is denoted by  $\kappa_{n}$.
The conventions of curvature tensors are $[\nabla _\rho ,\nabla_\sigma]V^\mu ={{\cal R}^\mu }_{\nu\rho\sigma}V^\nu$ 
and ${\cal R}_{\mu \nu }={{\cal R}^\rho }_{\mu \rho \nu }$ and the electromagnetic field strength is given by $F_{\mu\nu}:=\nabla_\mu A_\nu-\nabla_\nu A_\mu$.

%======================================%
%<<<<<<<<<<<< SECTION I  >>>>>>>>>>>>>>%
%======================================%
\section{Solution-generating transformations}
In the present paper, we consider the $n(\ge 4)$-dimensional Einstein-Maxwell system with a massless scalar field $\phi$, of which action is given by 
\begin{align}
\label{action}
S[g_{\mu\nu}, A_{\mu},\phi]=&\int \D^nx\sqrt{-g}\biggl(\frac{1}{2\kappa_{n}}{\cal R}-\frac{1}{4}F_{\mu\nu}F^{\mu\nu}-\frac12({\nabla} \phi)^2 \biggl). 
\end{align}
The field equations in this system are 
\begin{align}
&G_{\mu\nu}=\kappa_{n}\left(T^{(\rm em)}_{\mu\nu} +T^{(\phi)}_{\mu\nu}\right), \label{EFE}\\
&\nabla_\nu F^{\mu\nu}=0, \qquad \dalm\phi=0, \label{em-kg}
\end{align}
where $G_{\mu\nu}:={\cal R}_{\mu\nu}-(1/2)g_{\mu\nu}{\cal R}$ is the Einstein tensor and the energy-momentum tensors for the Maxwell field and the massless Klein-Gordon field are respectively given by 
\begin{align}
T^{(\rm em)}_{\mu\nu}:=&F_{\mu\rho}F_\nu^{~\rho}-\frac 14 g_{\mu\nu}F_{\rho\sigma}F^{\rho\sigma}, \label{Tab-Max}\\
T^{(\phi)}_{\mu\nu}: =&(\nabla_\mu \phi)(\nabla_\nu \phi)-\frac12 g_{\mu\nu} (\nabla\phi)^2.\label{Tab-scalar}
\end{align}

\subsection{Higher-dimensional Buchdahl and JRW transformations}
Now let us consider the most general $n$-dimensional static spacetimes ${\ma M}^n \approx \mathbb{R}\times M^{n-1}$ in the following form of the metric:
\begin{align}
\D s^2=&-\Omega(y)^{-2}\D t^2+\Omega(y)^{2/(n-3)}{\bar g}_{ij}(y)\D y^i\D y^j, \label{eq:structure-100}
\end{align}
where the indices $i$ and $j$ run from $1$ to $n-1$ and ${\bar g}_{ij}$ and $\Omega$ are an arbitrary Euclidean metric and a scalar function on $M^{n-1}$, respectively. 
It is noted that our metric ansatz (\ref{eq:structure-100}) is different from the one in the Tangen's formulation~\cite{tangen2007}.
In the following, we assume $\phi=\phi(y)$ and $A_\mu=A_t(y)\delta^t_\mu$ and the covariant derivative with respect to ${\bar g}_{ij}$ is introduced such that ${\bar D}_k{\bar g}_{ij}=0$.

Then, the basic equations (\ref{EFE}) and (\ref{em-kg}) reduce to
\begin{align}
&{}^{(n-1)}{\bar {\cal R}}_{ij}-\frac{n-2}{n-3}({\bar D}_i\ln\Omega)({\bar D}_j\ln\Omega) =\kappa_n({\bar D}_i \phi)({\bar D}_j \phi)-\kappa_n\Omega^2F_{it}F_{jt},\label{ein-eq1}\\
&-\frac{n-2}{n-3}{\bar D}^2\ln\Omega=\kappa_n\Omega^2{\bar g}^{kl}F_{tk}F_{tl}, \label{ein-eq3} \\
&{\bar D}^2\phi=0,\qquad \partial_i(\sqrt{\det{\bar g}}\Omega^2{\bar g}^{ij}F_{tj})=0,\label{matter-eq}
\end{align}
where ${\bar D}^2:={\bar g}^{ij}{\bar D}_i{\bar D}_j$ and a superscript $(n-1)$ implies a geometrical quantity constructed from ${\bar g}_{ij}$. (See Appendix~\ref{App:derivation} for derivation.) 
Equations (\ref{ein-eq1}) and (\ref{ein-eq3}) give the following auxiliary equation:
\begin{align}
{}^{(n-1)}{\bar {\cal R}}-\frac{n-2}{n-3}\left\{{\bar D}^2\ln\Omega+({\bar D}\ln\Omega)^2\right\}=\kappa_n({\bar D}\phi)^2.\label{ein-eq2}
\end{align}
It is observed that the basic equations \eqref{ein-eq1}--\eqref{matter-eq} has the symmetry
\begin{align}
\Omega \to \zeta \Omega, \qquad A_t \to \zeta^{-1}A_t,\label{symmetry}
\end{align}
where $\zeta$ is a non-zero real constant.

Now we are ready to formulate solution-generating transformations.
The following formulation of the $n$-dimensional generalization of the Buchdahl transformation is much simpler than the one in~\cite{tangen2007}.
%----------------------- lemma ------------------------------%
\begin{Prop}
\label{prop:Buchdahl}
Suppose that the following static metric in $n(\ge 4)$ dimensions
\begin{align}
\D s^2=-e^{2V}\D t^2+e^{-2V/(n-3)}{\bar g}_{ij}\D y^i\D y^j
\end{align}
solves the field equations (\ref{EFE}) and (\ref{em-kg}) with a constant scalar field and a trivial Maxwell field, where $V$ and ${\bar g}_{ij}$ are functions of $y^i$.
Then, the following static metric and scalar field solve the field equations (\ref{EFE}) and (\ref{em-kg}) with a trivial Maxwell field:
\begin{align}
\D s^2=&-e^{2U}\D t^2+e^{-2U/(n-3)}{\bar g}_{ij}\D y^i\D y^j,\\
U:=&\alpha V, \qquad \phi=\pm\sqrt{\frac{(n-2)(1-\alpha^2)}{\kappa_ {n}(n-3)}}V+\phi_0,\label{Buchdahl-scalar}
\end{align}
where $\alpha$ and $\phi_0$ are constants.
\end{Prop}
{\it Proof}. 
By Eqs.~(\ref{ein-eq1})--(\ref{matter-eq}), the field equations for the set $({\bar g}_{ij},\Omega,\phi,F_{ti})=({\bar g}_{ij},e^{-U},\phi,0)$ are given by 
\begin{align}
&{}^{(n-1)}{\bar {\cal R}}_{ij}-\frac{n-2}{n-3}({\bar D}_iU)({\bar D}_jU) =\kappa_n({\bar D}_i \phi)({\bar D}_j \phi),\label{set2-1}\\
&{\bar D}^2U=0,\qquad {\bar D}^2\phi=0. \label{set2-3}
\end{align}
Substituting Eq.~(\ref{Buchdahl-scalar}) into the above equations, we obtain 
\begin{align}
&{}^{(n-1)}{\bar {\cal R}}_{ij}-\frac{n-2}{n-3}({\bar D}_iV)({\bar D}_jV) =0,\label{set3-1}\\
&{\bar D}^2V=0.\label{set3-3}
\end{align}
These are the field equations for the set $({\bar g}_{ij},\Omega,\phi,F_{ti})=({\bar g}_{ij},e^{-V},\Phi_0,0)$, where $\Phi_0$ is a constant.
\qed
%----------------------- lemma ------------------------------%

Next, we formulate an $n$-dimensional generalization of the JRW transformation in a similar manner.
%----------------------- lemma ------------------------------%
\begin{Prop}
\label{prop:JRW}
Suppose that the following static metric in $n(\ge 4)$ dimensions
\begin{align}
\D s^2=-e^{2U}\D t^2+e^{-2U/(n-3)}{\bar g}_{ij}\D y^i\D y^j \label{metric-JRW-trans}
\end{align}
and a scalar field $\phi$ solve the field equations (\ref{EFE}) and (\ref{em-kg}) with a trivial Maxwell field, where $U$, ${\bar g}_{ij}$, and $\phi$ are functions of $y^i$.
Then, the following static metric
\begin{align}
\D s^2=&-e^{2W}\D t^2+e^{-2W/(n-3)}{\bar g}_{ij}\D y^i\D y^j,\\
W:=&-\ln|\sinh (U-U_0)| \label{JRW-W}
\end{align}
and the same form of the scalar field $\phi$ solve the field equations (\ref{EFE}) and (\ref{em-kg}) with the following Maxwell field:
\begin{align}
F_{ti}=&\pm \sqrt{\frac{n-2}{(n-3)\kappa_n}}\frac{{\bar D}_iU}{\sinh^2 (U-U_0)},\label{JRW-Max}\\
A_t=&\pm\sqrt{\frac{n-2}{(n-3)\kappa_n}}\frac{1}{\tanh (U-U_0)}+A_0, 
\end{align}
where $U_0$ and $A_0$ are constants.
\end{Prop}
{\it Proof}. 
By Eqs.~(\ref{ein-eq1})--(\ref{matter-eq}), the field equations for the set $({\bar g}_{ij},\Omega,\phi,F_{ti})=({\bar g}_{ij},e^{-W},\phi,F_{ti})$ are given by 
\begin{align}
&{}^{(n-1)}{\bar {\cal R}}_{ij}-\frac{n-2}{n-3}({\bar D}_iW)({\bar D}_jW) =\kappa_n({\bar D}_i \phi)({\bar D}_j \phi)-\kappa_n e^{-2W}F_{it}F_{jt},\label{set1-1}\\
&\frac{n-2}{n-3}{\bar D}^2W=\kappa_ne^{-2W}{\bar g}^{kl}F_{tk}F_{tl}, \label{set1-3}\\
&{\bar D}^2\phi=0,\qquad \partial_i(\sqrt{\det{\bar g}}e^{-2W}{\bar g}^{ij}F_{tj})=0.\label{set1-4}
\end{align}
Substituting Eqs.~(\ref{JRW-W}) and (\ref{JRW-Max}) into Eqs.~(\ref{set1-1})--(\ref{set1-4}), we obtain the field equations (\ref{set2-1}) and (\ref{set2-3}) for the set $({\bar g}_{ij},\Omega,\phi,F_{ti})=({\bar g}_{ij},e^{-U},\phi,0)$.
\qed
%----------------------- lemma ------------------------------%

We note that the constant $U_0$ in Eqs.~\eqref{JRW-W} and \eqref{JRW-Max} does not appear in the original formulation in four dimensions~\cite{jrw1969}.
Actually, this constant is crucial for providing a new integration constant, as shown in the next section.

\subsection{Derivations in the nonlinear sigma model approach}

In this subsection, we present alternative derivations of the higher-dimensional Buchdahl and JRW transformations based on the reduced action in the form of a nonlinear sigma model.

Using the decomposition of the Ricci scalar ${}^{(n)}{\cal R}$ for static spacetimes (\ref{eq:structure-100}) shown in Appendix \ref{App:derivation} together with the assumptions $\phi=\phi(y)$ and $A_\mu=A_t(y)\delta^t_\mu$, we obtain the following reduced action for static solutions:
\begin{align}
\label{action-reduce}
S_{n-1}=&\int \D^{n-1}x\sqrt{-{\bar g}}\biggl\{\frac{1}{2\kappa_{n}}\biggl({}^{(n-1)}{\bar {\cal R}}-\frac{2}{n-3}{\bar D}^2\ln\Omega-\frac{n-2}{n-3}({\bar D}\ln\Omega)^2\biggl) \nonumber \\
&+\frac{1}{2}\Omega^{2}({\bar D} A_t)^2-\frac12({\bar D} \phi)^2 \biggl\}. 
\end{align}
Omitting the total derivative term with ${\bar D}^2\ln\Omega$, we can write the above reduced action in the following form of a nonlinear sigma model $\psi^{(a)}~(a=1,2,3)$:
\begin{align}
\label{action-reduce2}
S_{n-1}=&\int \D^{n-1}x\sqrt{-{\bar g}}\biggl\{\frac{1}{2\kappa_{n}}{}^{(n-1)}{\bar {\cal R}}-\frac12{\cal G}_{(a)(b)}({\bar D}_i\psi^{(a)})({\bar D}^i\psi^{(b)})\biggl\},\\
\psi^{(1)}:=&-\ln\Omega,\qquad \psi^{(2)}:=\sqrt{\frac{(n-3)\kappa_n}{n-2}}A_t,\qquad \psi^{(3)}:=\sqrt{\frac{(n-3)\kappa_n}{n-2}}\phi,
\end{align}
where the metric ${\cal G}_{(a)(b)}$ in the three-dimensional target space is given by
\begin{align}
\D{\cal S}^2=&{\cal G}_{(a)(b)}\D\psi^{(a)}\D\psi^{(b)} \nonumber \\
=&\frac{n-2}{(n-3)\kappa_n}\left((\D {\psi}^{(1)})^2-e^{-2{\psi}^{(1)}}(\D{\psi}^{(2)})^2+(\D{\psi}^{(3)})^2\right).
\end{align}
A transformation ${\bar\psi}^{(a)}={\bar\psi}^{(a)}(\psi)~(a=1,2,3)$ keeping $\D {\cal S}^2$ invariant corresponds to a solution-generating transformation.

In the absence of a Maxwell field, the target space is a simple two-dimensional flat space:
\begin{align}
\D{\cal S}^2=\frac{n-2}{(n-3)\kappa_n}\left((\D {\psi}^{(1)})^2+(\D{\psi}^{(3)})^2\right).
\end{align}
In this case of linear sigma model, a new solution is generated by a linear transformation characterized by a $2\times 2$ matrix such that
\begin{equation}
\begin{pmatrix}
{\bar\psi}^{(1)}\\
{\bar\psi}^{(3)}
\end{pmatrix}
=
\begin{pmatrix}
\cos\theta & -\sin\theta\\
\sin\theta  &\cos\theta
\end{pmatrix}
\begin{pmatrix}
{\psi}^{(1)}\\
{\psi}^{(3)}
\end{pmatrix}
,
\end{equation}
where $\theta\in  \mathbb{C}$  is a parameter of the transformation\footnote{The higher-dimensional JNW solution has been derived by this method in~\cite{as2010}.}.
The Buchdahl transformation\eqref{Buchdahl-scalar} (with $\phi_0=0$) is a special case with ${\psi}^{(3)}=0$ and a reparametrization $\alpha=\cos\theta$:
\begin{equation}
\begin{pmatrix}
{\bar\psi}^{(1)}\\
{\bar\psi}^{(3)}
\end{pmatrix}
=
\begin{pmatrix}
\alpha & \mp \sqrt{1-\alpha^2}\\
\pm\sqrt{1-\alpha^2}  &\alpha
\end{pmatrix}
\begin{pmatrix}
{\psi}^{(1)}\\
0
\end{pmatrix}
.
\end{equation}
A real scalar field is obtained if $\theta$ is real and a ghost scalar field is given when $\theta$ is chosen to be purely imaginary.

A solution-generating transformation ${\bar\psi}^{(a)}={\bar\psi}^{(a)}(\psi)~(a=1,2,3)$ becomes nonlinear in the presence of a Maxwell field.
In order to find such a transformation, we rewrite the reduce action (\ref{action-reduce2}) in the following form~\cite{gr1998,Yazadjiev2005}:
\begin{align}
\label{action-reduce3}
S_{n-1}=&\int \D^{n-1}x\sqrt{-{\bar g}}\biggl\{\frac{1}{2\kappa_{n}}{}^{(n-1)}{\bar {\cal R}}+\frac{n-2}{4(n-3)\kappa_n}{\rm Tr}\left(({\bar D}_i P){\bar D}^i(P^{-1})\right)\biggl\},
\end{align}
where the matrix $P$ is defined by
\begin{align}
P:=e^{-{\psi}^{(1)}-{\psi}^{(3)}}
\begin{pmatrix}
e^{2{\psi}^{(1)}}-({\psi}^{(2)})^2 & -{\psi}^{(2)} \\
-{\psi}^{(2)} & -1
\end{pmatrix}
.
\end{align}
A key fact is that 
\begin{align}
{\rm Tr}\left((D_i P')D^i({P'}^{-1})\right)={\rm Tr}\left((D_i P)D^i(P^{-1})\right)
\end{align}
holds for $P':=GPG^T$ with any $2\times 2$ constant matrix $G\in {\rm GL}(2,\mathbb{C})$.
With the general representation of $G$ such that
\begin{equation}
\label{matrix-G}
G=
\begin{pmatrix}
a & b \\
c & d 
\end{pmatrix}
,
\end{equation}
the relation $P'=GPG^T$ gives the following three equations:
\begin{align}
e^{-{\bar\psi}^{(1)}}\left(e^{2{\bar\psi}^{(1)}}-({\bar\psi}^{(2)})^2\right)=&e^{-{\psi}^{(1)}}\left\{a^2e^{2{\psi}^{(1)}}-(a{\psi}^{(2)}+b)^2\right\},\label{rel1}\\
-e^{-{\bar\psi}^{(1)}}{\bar\psi}^{(2)}=&e^{-{\psi}^{(1)}}\left\{ace^{2{\psi}^{(1)}}-ac({\psi}^{(2)})^2-(ad+bc){\psi}^{(2)}-bd\right\},\label{rel2}\\
-e^{-{\bar\psi}^{(1)}}=&e^{-{\psi}^{(1)}}\left\{c^2e^{2{\psi}^{(1)}}-(c{\psi}^{(2)}+d)^2\right\}.\label{rel3}
\end{align}
Substituting Eqs.~(\ref{rel2}) and (\ref{rel3}) into Eq.~(\ref{rel1}), we obtain
\begin{align}
(ad-bc)^2=1
\end{align}
and hence $\det G=ad-bc=\pm1$ is required for consistency.
Thus, a matrix (\ref{matrix-G}) whose determinant is $\pm1$ defines a solution-generating transformation from $\{{\psi}^{(1)},{\psi}^{(2)},{\psi}^{(3)}\}$ to $\{{\bar\psi}^{(1)},{\bar\psi}^{(2)},{\psi}^{(3)}\}$ by Eqs.~(\ref{rel2}) and (\ref{rel3}), which are explicitly written as
\begin{align}
e^{-{\bar\psi}^{(1)}}=&-e^{-{\psi}^{(1)}}\left\{c^2e^{2{\psi}^{(1)}}-(c{\psi}^{(2)}+d)^2\right\},\label{rel3-2}\\
{\bar\psi}^{(2)}=&\left\{c^2e^{2{\psi}^{(1)}}-(c{\psi}^{(2)}+d)^2\right\}^{-1}\left\{ace^{2{\psi}^{(1)}}-ac({\psi}^{(2)})^2-(ad+bc){\psi}^{(2)}-bd\right\}. \label{rel2-2}
\end{align}
While $\det G=1$ implies $G\in {\rm SL}(2,\mathbb{C})$, performing a transformation with $\det G=-1$ twice gives a single transformation whose determinant equals $1$.

Since the scalar field is unchanged in the resulting solution, the Buchdahl transformation is not contained by this class of transformations.
On the other hand, the JRW transformation is contained in this class, in which the seed solution in the JRW transformation is neutral and hence ${\psi}^{(2)}=0$.
In this case, the new solution is given by 
\begin{align}
e^{-{\bar\psi}^{(1)}}=&-\left(c^2e^{{\psi}^{(1)}}-d^2e^{-{\psi}^{(1)}}\right),\label{rel3-3}\\
{\bar\psi}^{(2)}=&\frac{ace^{{\psi}^{(1)}}-bde^{-{\psi}^{(1)}}}{c^2e^{{\psi}^{(1)}}-d^2e^{-{\psi}^{(1)}}}. \label{rel2-3}
\end{align}
Note that the transformation matrix with $cd=0$ does not generate a new solution. Moreover, taken into account the symmetry \eqref{symmetry} of the basic equations, we can choose $c^2 d^2=1/4$ without loss of generality. Thus, the components of the matrix $G$ for the JRW transformation (with $A_0=0$) satisfy
\begin{align}
c^2=\pm\frac12e^{-U_0},\quad d^2=\pm\frac12e^{U_0}, \quad ac=\frac12e^{-U_0},\quad bd=-\frac12e^{U_0}, \label{abcd}
\end{align}
with which Eqs.~(\ref{rel3-3}) and (\ref{rel2-3}) respectively give
\begin{align}
e^{-{\bar\psi}^{(1)}}=&\mp\frac12(e^{{\psi}^{(1)}-U_0}-e^{-{\psi}^{(1)}+U_0}) =|\sinh({\psi}^{(1)}-U_0)|,\\
{\bar\psi}^{(2)}=&\pm\frac{e^{{\psi}^{(1)}-U_0}+ e^{-{\psi}^{(1)}+U_0}}{e^{{\psi}^{(1)}-U_0}-e^{-{\psi}^{(1)}+U_0}}=\pm\frac{1}{\tanh({\psi}^{(1)}-U_0)}.
\end{align}
The upper (lower) signs in Eq.~(\ref{abcd}) correspond to the case of $\sinh(U-U_0)\le(\ge)0$. 
The condition $(ad-bc)^2=1$ certainly holds for a set $\{a,b,c,d\}$ satisfying Eq.~(\ref{abcd}), where the upper (lower) signs make all the components $a$, $b$, $c$, and $d$ real (pure imaginary).
We note that the constants $a$ and $b$ correspond to pure gauge in the resulting gauge potential ${\bar\psi}^{(2)}$.
This is seen by rewriting Eq.~(\ref{rel2-3}) as
\begin{align}
{\bar\psi}^{(2)}=\frac{cd(ad-bc)(e^{{\psi}^{(1)}}-e^{-{\psi}^{(1)}})}{(d^2-c^2)(c^2e^{{\psi}^{(1)}}-d^2e^{-{\psi}^{(1)}})}-\frac{ac-bd}{d^2-c^2}, \end{align}
where $(ad-bc)^2=1$ is required.
Since $a$ and $b$ correspond to pure gauge and one can set $c^2 d^2=1/4$ without loss of generality, the transformation (\ref{rel3-3}) and (\ref{rel2-3}) is uniparametric and identical to the JRW transformation (\ref{JRW-W}) and (\ref{JRW-Max}).

In~\cite{Yazadjiev2005}, the present approach has been adopted to establish a solution-generating technique in the arbitrary-dimensional Einstein-Maxwell-dilaton system.  
There the author presented two concrete transformations to obtain charged solutions from a vacuum solution.
Since the higher-dimensional JRW transformation, given by  \eqref{JRW-W} and \eqref{JRW-Max}, generates a charged solution from a neutral non-vacuum solution, it is not realized in the limit of the vanishing dilaton coupling $\alpha\to 0$ in the transformations considered in~\cite{Yazadjiev2005}. 
(See~\cite{lim2017} for a similar but different approach to obtain static solutions in the arbitrary-dimensional Einstein-Maxwell-dilaton system with or without a cosmological constant.)

%======================================%
%<<<<<<<<<<<< SECTION I  >>>>>>>>>>>>>>%
%======================================%
\section{Applications}
\label{sec:application}
We have established in the previous section the arbitrary $n(\ge 4)$-dimensional generalizations of the Buchdahl and JRW transformations.
In this section, we adopt these transformations to some specific solutions.

\subsection{Static solutions with Einstein base manifold}
First let us consider static solutions with possible additional symmetries provided by a $(n-2)$-dimensional Einstein space.
The vacuum seed solution we consider is the following topological generalization of the Schwarzschild-Tangherlini solution:
\begin{align}
\D s^2=&-f(r) \D { t}^2+f(r)^{-1}\D r^2+r^2\gamma_{ab}(z)\D z^a\D z^b,\\
f(r)=&k-\frac{\mu}{r^{n-3}},
\end{align}
where $\mu$ is a constant.
Here $\gamma_{ab}(z)$ is the metric on the $(n-2)$-dimensional Einstein space $K^{n-2}$, whose Ricci tensor is given by ${}^{(n-2)}{\cal R}_{ab}=k(n-3)\gamma_{ab}$, where $k =1,0,-1$.
Adopting the $n(\ge 4)$-dimensional Buchdahl transformation to the above solution, we obtain the following generalized JNW solution~\cite{mm2016}\footnote{An $(n-2)$-dimensional sphere was previously considered as $K^{n-2}$ in \cite{JNWhigher}.}:
\begin{align}
\D s^2=&-f(r)^\alpha \D { t}^2+f(r)^{-\alpha/(n-3)}\biggl(f(r)^{-(n-4)/(n-3)}\D r^2+r^2f(r)^{1/(n-3)}\gamma_{ab}(z)\D z^a\D z^b\biggl),\label{JNW-higher-200}\\
\phi =&\phi_0\pm\sqrt{\frac{(n-2)(1-\alpha^2)}{4\kappa_ {n}(n-3)}}\ln f(r), \qquad f(r)=k-\frac{\mu}{r^{n-3}}. \label{JNW-higher-scalar-200}
\end{align}

In the classification of static solutions with an Einstein base manifold in the present system (\ref{action}) performed in~\cite{mm2016}, we have shown that the generalized JNW solution (\ref{JNW-higher-200}) is the unique neutral solution with a nontrivial scalar field.
Then, which charged solution in~\cite{mm2016} is obtained by the $n(\ge 4)$-dimensional JRW transformation from the generalized JNW solution?

Indeed, the resulting solution after the JRW transformation is given by 
\begin{align}
\D s^2=&-\biggl(\frac{e^{-U_0}f^{\alpha/2}-e^{U_0}f^{-\alpha/2}}{2}\biggl)^{-2}\D { t}^2+\biggl(\frac{e^{-U_0}f^{\alpha/2}-e^{U_0}f^{-\alpha/2}}{2}\biggl)^{2/(n-3)} \nonumber \\
&\times \biggl(f^{-(n-4)/(n-3)}\D r^2+r^2f^{1/(n-3)}\gamma_{ab}(z)\D z^a\D z^b\biggl),\label{JNW-higher-charged}\\
F_{tr}=&\pm \sqrt{\frac{(n-2)(n-3)}{\kappa_n}}\frac{2 \alpha\mu}{f r^{n-2}(e^{-U_0}f^{\alpha/2}-e^{U_0}f^{-\alpha/2})^{2}}, \label{electf}
\end{align}  
where the scalar field remains in the same form (\ref{JNW-higher-scalar-200}).
We note that the $\pm$ signs in $\phi$ and $F_{tr}$ are independent in the present subsection.
We are going to show that this is the type-I solution for $k=1,-1$ and the type-VI${}_0$ solution for $k=0$ in the classification~\cite{mm2016}.

The metric of the type-I solution for $k=1,-1$ is given by 
\begin{align}
\D s^2=&-F(x)^{-2}\D t^2+F(x)^{2/(n-3)}G(x)^{-(n-4)/(n-3)}\biggl(\D x^2+G(x)\gamma_{ab}(z)\D z^a\D z^b\biggl), \label{metric-base} \\
F(x)=&A\biggl(\varepsilon\frac{x-x_0}{x+x_0}\biggl)^{\alpha/2}+B\biggl(\varepsilon\frac{x-x_0}{x+x_0}\biggl)^{-\alpha/2}, \qquad G(x)=k(n-3)^2(x^2-x_0^2),
\end{align}
where $x_0\ne 0$ is required and $\varepsilon=\pm1$ is put in order for the interior of the bracket to be non-negative\footnote{Without loss of generality, we have applied the change of coordinate ${\bar x}:=x-(a+b)/2$ to the original solution~\cite{mm2016} and defined $x_0:=(a-b)/2$. We have then omitted the bars for simplicity.}.
The constants $A$ and $B$ are arbitrary constants not simultaneously null\footnote{Both in the type-I and type-VI${}_0$ solutions, one can set $A$ or $B$ be any non-zero value without loss of generality if it is non-zero by scaling transformations of $t$ and $x$, however we keep both of them to be arbitrary.}.  
The scalar field of the type-I solution is 
\begin{align}
\phi(x)=\phi_0\pm\sqrt{ \frac{(n-2)(1-\alpha^2)}{4\kappa_{n} (n-3)}}\ln\biggl(\varepsilon\frac{x-x_0}{x+x_0}\biggl),\label{TypeI-phi}
\end{align}
and the  electric field is given by
\begin{align} \label{electfold}
F_{tx}=-\frac{q}{F^{2}G}.
\end{align}
The constant $\alpha$ and the electric charge $q$ are related with the other integration constants as
\begin{align} \label{q}
\kappa_{n} q^2=-4 (n-2)(n-3)^3 A B \alpha^2 x_0^2. 
\end{align}
Now let us consider the coordinate transformation
\begin{align} \label{ct}
x=\frac{1+k f(r)}{1-k f(r)} x_0.
\end{align}
and rewrite $x_0$ as $x_0=\mu/[2(n-3)]$, which give
\begin{align}
\frac{x-x_0}{x+x_0}=k f(r)=1-\frac{k \mu}{r^{n-3}}.
\end{align}
After this transformation, the type-I solution (\ref{metric-base})--(\ref{electfold}) becomes 
\begin{align}
\D s^2=&-\biggl\{A(\varepsilon k f)^{\alpha/2}+B (\varepsilon k f)^{-\alpha/2}\biggl\}^{-2}\D t^2+\biggl\{A(\varepsilon k f)^{\alpha/2}+B(\varepsilon k f)^{-\alpha/2}\biggl\}^{2/(n-3)} \nonumber \\
&\times\biggl(f^{-(n-4)/(n-3)}\D r^2+r^2f^{1/(n-3)}\gamma_{ab}(z)\D z^a\D z^b\biggl)
\end{align}
with
\begin{align}
\phi(r)=&\phi_0\pm\sqrt{ \frac{(n-2)(1-\alpha^2)}{4\kappa_{n} (n-3)}}\ln(\varepsilon k f),\label{TypeI-phi-2} \\
F_{tr}=&\pm\sqrt{-\frac{(n-2)(n-3) A B \alpha^2 \mu^2}{\kappa_{n}}}\frac{1}{f r^{n-2}[A(\varepsilon k f)^{\alpha/2}+B (\varepsilon k f)^{-\alpha/2}]^2}, \label{electf-2}
\end{align}
where we used Eq.~(\ref{q}).
Finally, setting $\varepsilon$ such that $\varepsilon k=1$ and identifying $A=\pm e^{-U_0}/2$ and $B=\mp e^{U_0}/2$, we recover the metric~(\ref{JNW-higher-charged}), the scalar field \eqref{JNW-higher-scalar-200}, and the Maxwell field \eqref{electf}.

On the other hand, the metric of the type-VI${}_0$ solution for $k=0$ is given by Eq.~(\ref{metric-base}) with 
\begin{align}
F(x)=& A(\varepsilon G_1 x)^{\alpha/2}+B(\varepsilon G_1 x)^{-\alpha/2}, \qquad G(x)=G_1 x,
\end{align}
where $G_1$ is a non-zero constant\footnote{We have performed the change of coordinate ${\bar x}:=x+ G_0/G_1$ in the original type-VI${}_0$ solution presented in \cite{mm2016} and then omitted the bars for simplicity.}. 
The scalar field in this solution is given by 
\begin{align}
\phi(x)={\bar \phi}_0\mp\sqrt{\frac{(n-2)(1-\alpha^2)}{4(n-3)\kappa_{n}}}\ln(\varepsilon G_1 x),\label{k=0}
\end{align}
where ${\bar \phi}_0$ is a constant, and the Maxwell field reads
\begin{align} \label{electfk0}
F_{tx}=-\frac{q}{G_1 x[A(\varepsilon G_1 x)^{\alpha/2}+B(\varepsilon G_1 x)^{-\alpha/2}]^2}.
\end{align}
The electric charge $q$ is related with the other integration constants as
\begin{align} \label{q2}
\kappa_{n} q^2 =-\frac{(n-2)\alpha^2G_1^2 A B}{(n-3)}. 
\end{align}
In what follows, we use $f(r)= -\mu/r^{n-3}$ and rewrite $G_1$ such that $G_1^2=(n-3)^2 \mu^{2}$. Then, the coordinate transformation
\begin{align}
G_1 x=\frac{G_1^2}{(n-3)^2f(r)}=\frac{\mu^2}{f(r)}=-\mu r^{n-3},
\end{align}
 allows us to write the type-VI${}_0$ solution as 
\begin{align}
\D s^2=&-\biggl\{B(\varepsilon\mu^{-2}f)^{\alpha/2}+A(\varepsilon\mu^{-2}f)^{-\alpha/2}\biggl\}^{-2}\D t^2+\biggl\{B(\varepsilon\mu^{-2}f)^{\alpha/2}+A(\varepsilon\mu^{-2}f)^{-\alpha/2}\biggl\}^{2/(n-3)} \nonumber \\
&\times \biggl(f^{-(n-4)/(n-3)}\D r^2+r^2f^{1/(n-3)}\gamma_{ab}(z)\D z^a\D z^b\biggl)
\end{align}
with
\begin{align}
\phi(r)=&{\bar \phi}_0\mp\sqrt{\frac{(n-2)(1-\alpha^2)}{4(n-3)\kappa_{n}}}\ln(\varepsilon \mu^2f^{-1}) \nonumber \\
=&\biggl({\bar \phi}_0\mp\sqrt{\frac{(n-2)(1-\alpha^2)}{4(n-3)\kappa_{n}}}\ln(\varepsilon \mu^2)\biggl)\pm\sqrt{\frac{(n-2)(1-\alpha^2)}{4(n-3)\kappa_{n}}}\ln f,\label{k=0-2}\\
F_{tr}=&\mp \sqrt{-\frac{(n-2)(n-3)\alpha^2A B}{\kappa_{n} }}\frac{1}{r[B(\varepsilon\mu^{-2}f)^{\alpha/2}+A(\varepsilon\mu^{-2}f)^{-\alpha/2}]^2},
\end{align}
where we used Eq.~(\ref{q2}) and the following relations holding for $k=0$:
\begin{align}
f^{-1/(n-3)}=\frac{r^2f^{1/(n-3)}}{(-\mu)^{2/(n-3)}},\qquad f^{(n-4)/(n-3)}=\frac{(-\mu)^{2(n-4)/(n-3)}}{r^{2(n-4)}f^{(n-4)/(n-3)}}.
\end{align}
Finally, identifying $A$ and $B$ such that
\begin{align}
A=\pm \frac{1}{2}(\varepsilon\mu^{-2})^{\alpha/2}e^{U_0},\qquad B=\mp \frac{1}{2}(\varepsilon\mu^{-2})^{-\alpha/2}e^{-U_0},
\end{align}  
we recover the metric~(\ref{JNW-higher-charged}), the scalar field \eqref{JNW-higher-scalar-200}, and the Maxwell field \eqref{electf} with $k=0$.

\subsection{Multi-center solutions}
Next, as another demonstration, we consider the case where ${\bar g}_{ij}(y)\D y^i\D y^j$ is flat:
\begin{align}
{\bar g}_{ij}(y)\D y^i\D y^j=\D y_1^2+\D y_2^2+\cdots\D y_{n-1}^2.
\end{align}
In this case, the field equations (\ref{ein-eq1})--(\ref{matter-eq}) with a constant scalar field admit the following multi-center solution:
\begin{align}
{\bar D}^2\Omega=0,\qquad F_{ti}=\pm\sqrt{\frac{n-2}{(n-3)\kappa_n}}\Omega^{-2}\partial_i\Omega.
\end{align}
The first equation shows that $\Omega$ is a harmonic function on the $(n-1)$-dimensional flat space and hence $\Omega$ can be expressed as a linear combination of the fundamental solution, namely
\begin{align}
\Omega=&\mu_0+\sum_{p=1}^N\frac{\mu_p}{r_p^{n-3}},\\
r_p:=&\sqrt{(y_1-c_{p(1)})^2+(y_2-c_{p(2)})^2+\cdots+(y_{n-1}-c_{p(n-1)})^2},
\end{align}
where $\mu_0$, $\mu_p$, $c_{p(1)},c_{p(2)},\cdots,c_{p(n-1)}$ are constants and $N$ is a positive integer.
This is the $n(\ge 4)$-dimensional generalization of the Majumdar-Papapetrou solution~\cite{majumdar1947,papapetrou1947,lz2005}.
The original Majumdar-Papapetrou solution represents a static configuration of multi-black holes with degenerate horizons, where $r_p=0$ defines the location of the Killing horizon of each black hole~\cite{hh1972}.
A remarkable feature of this solution is that, while the spacetime are analytic at the horizons in four dimensions~\cite{hh1972}, the differentiability of the spacetime at the horizons becomes lower in higher dimensions~\cite{cr2007}.
Indeed, in six or higher dimensions, there appear parallelly propagated (p.p.) curvature singularities~\cite{Hawking:1973uf} at the location of the horizons and then the solution does not represent a configuration of multi-black holes any more.

On the other hand, in the absence of a Maxwell field, the field equations (\ref{ein-eq1})--(\ref{matter-eq}) admit the following multi-center solution:
\begin{align}
{\bar D}^2\ln\Omega=0,\qquad \phi=\phi_0\pm \sqrt{-\frac{n-2}{(n-3)\kappa_n}}\ln\Omega, \label{multi-sol}
\end{align}
where $\phi_0$ is a constant.
These equations show that both $\ln \Omega$ and $\phi$ are harmonic functions and $\phi-\phi_0$ is pure imaginary, namely $\phi$ is a ghost scalar field.
The explicit form of $\Omega$ is given by 
\begin{align}
\ln \Omega=&\nu_0+\sum_{p=1}^N\frac{\nu_p}{r_p^{n-3}},\label{lnOmega}\\
r_p:=&\sqrt{(y_1-d_{p(1)})^2+(y_2-d_{p(2)})^2+\cdots+(y_{n-1}-d_{p(n-1)})^2},
\end{align}
where $\nu_0$, $\nu_p, d_{p(1)},d_{p(2)},\cdots,d_{p(n-1)}$ are constants.
This is a generalization of the four-dimensional solution obtained by Gibbons~\cite{gibbons2003}\footnote{This solution has also been obtained in the axisymmetric case~\cite{ak1993}, but the authors did not point out that the scalar field is ghost.}.

This spacetime with $\nu_p=0$ for all $p$ is Minkowski.
For $\nu_p\ne 0$ for some $p$, the spacetime is asymptotically flat for $r:=\sqrt{y_1^2+y_2^2+\cdots+y_{n-1}^2}\to \infty$, namely at spacelike infinity.
On the other hand, $\Omega\to +\infty(+0)$ holds in the limit of $r_p\to 0$ for $\nu_p>(<)0$.
The Ricci scalar of this spacetime is computed to give
\begin{align}
{}^{(n)}{\cal R}=&-\frac{n-2}{n-3}\Omega^{-2/(n-3)}({\bar D}\ln\Omega)^2\nonumber \\
=&-(n-2)(n-3)\Omega^{-2/(n-3)}\delta^{ij}\sum_{p=1}^N\frac{\nu_p}{r_p^{n-1}}(y_i-d_{p(i)})\sum_{q=1}^N\frac{\nu_q}{r_q^{n-1}}(y_j-d_{q(j)}). \label{R-multi}
\end{align} 
In the particular case with $N=2$, the above expression reduces to
\begin{align}
{}^{(n)}{\cal R}=&-(n-2)(n-3)\Omega^{-2/(n-3)}\biggl\{\frac{\nu_1^2}{r_1^{2(n-2)}}+\frac{2\nu_1\nu_2}{r_1^{n-1}r_2^{n-1}}\delta^{ij}(y_i-d_{1(i)})(y_j-d_{2(j)}) +\frac{\nu_2^2}{r_2^{2(n-2)}}\biggl\}.
\end{align} 
Equation~(\ref{R-multi}) shows that there are naked singularities located at $r_p=0$ for $\nu_p<0$.
In contrast, ${}^{(n)}{\cal R}\to 0$ holds in the limit of $r_p\to 0$ for $\nu_p>0$.
According to the studies in the four-dimensional axisymmetric case~\cite{eks2016,gv2017}, $r_p=0$ with $\nu_p>0$ could be a wormhole throat, but a careful analysis of the spacetime is required at $r_p=0$ to provide its physical interpretation correctly.

Now let us obtain a multi-center solution both with a Maxwell field and a ghost scalar field by the $n$-dimensional JRW transformation from the solution (\ref{multi-sol}), where we identify $U=-\ln \Omega$ and then $e^{2W}=1/\sinh^{2}(\ln\Omega)$.
The resulting charged solution is given by 
\begin{align}
\D s^2=&-\frac{\D t^2}{\sinh^{2}(\ln\Omega)}+[\sinh^{2}(\ln\Omega)]^{1/(n-3)} (\D y_1^2+\D y_2^2+\cdots\D y_{n-1}^2), \label{new-sol}\\
F_{ti}=&\mp \sqrt{\frac{n-2}{(n-3)\kappa_n}}\frac{\partial_i(\ln\Omega)}{\sinh^2(\ln\Omega)},\qquad \phi=\phi_0\pm \sqrt{-\frac{n-2}{(n-3)\kappa_n}}\ln\Omega,
\end{align}
where $\ln\Omega$ is given by Eq.~(\ref{lnOmega}).
In~\cite{fz2012}, a static configuration of a massless scalar field or a Maxwell field is studied in the $n(\ge 4)$-dimensional Majumdar-Papapetrou background spacetime.
In contrast, our solution (\ref{new-sol}) is an exact solution taking the backreaction into the account completely.

\subsection{Cylindrically symmetric solutions in four dimensions}
Lastly, we present an application of the Buchdahl and JRW transformations to cylindrically symmetric spacetimes in four dimensions.
As a seed solution, we consider the Levi-Civita vacuum solution in the following coordinates:
\begin{equation}
\D s^2 =-{\rho}^{4\sigma}\D {t}^2 +{\rho}^{-4\sigma}\left\{{\rho}^{8\sigma^2}(\D{\rho}^2+\D {z}^2) +{C}^2{\rho}^{2}\D\phi^2\right\},\label{LeviCivita2}
\end{equation}
where $\sigma=0$ corresponds to Minkowski. 
The parameter $\sigma$ is interpreted as the mass per unit length of the source located along the axis $\rho=0$ and $C$ is the conicity parameter.
(See section 22.2 in~\cite{Stephani:2003} and section 10.2 in~\cite{GriffithsPodolsky:2009}.)

By the Buchdahl transformation (\ref{Buchdahl-scalar}) from the Levi-Civita solution (\ref{LeviCivita2}), one obtains
\begin{align}
\D s^2 =&-{\rho}^{4\alpha\sigma}\D {t}^2 +{\rho}^{-4\alpha\sigma}\left\{{\rho}^{8\sigma^2}(\D{\rho}^2+\D {z}^2) +{C}^2{\rho}^{2}\D\phi^2\right\},\label{LeviCivita3}\\
\phi=&\pm2\sigma\sqrt{\frac{2(1-\alpha^2)}{\kappa_ {4}}}\ln{\rho}+\phi_0.\label{LeviCivita3-scalar}
\end{align}
This Levi-Civita solution with a massless scalar hair was obtained in~\cite{em2015}.
Then, by the JRW transformation (\ref{JRW-W}) and (\ref{JRW-Max}) from the above solution, we obtain
\begin{align}
\D s^2 =&-4\left({\rho}^{2\alpha\sigma} e^{-U_0}-{\rho}^{-2\alpha\sigma} e^{U_0}\right)^{-2}\D {t}^2 \nonumber \\
&+\frac14\left({\rho}^{2\alpha\sigma} e^{-U_0}-{\rho}^{-2\alpha\sigma} e^{U_0}\right)^2\left\{{\rho}^{8\sigma^2}(\D{\rho}^2+\D {z}^2) +{C}^2{\rho}^{2}\D\phi^2\right\},\label{LeviCivita4}\\
F_{t{\rho}}=&\pm \sqrt{\frac{2}{\kappa_4}}\frac{8\alpha\sigma}{{\rho}({\rho}^{2\alpha\sigma} e^{-U_0}-{\rho}^{-2\alpha\sigma} e^{U_0})^{2}}
\end{align}
with the same form of the scalar field (\ref{LeviCivita3-scalar}), where the signs of $\phi$ and $F_{t{\rho}}$ are independent.
For $\alpha^2=1$, the scalar field becomes trivial and this charged Levi-Civita solution with a massless scalar hair reduces to the Raychaudhuri solution in the Einstein-Maxwell system~\cite{Raychaudhuri1960}. (See section 22.2 in~\cite{Stephani:2003}.)

%======================================%
%<<<<<<<<<<<< SECTION I  >>>>>>>>>>>>>>%
%======================================%
\section{Summary and future prospects}
In the present paper, we have presented higher-dimensional generalizations of the Buchdahl and JRW transformations which generate static solutions in the Einstein-Maxwell system with a massless scalar field.
While the former adds a nontrivial scalar field to a vacuum solution, the latter generates a charged solution from a neutral one with the same form of a scalar field.
Our formulation of the Buchdahl transformation is simpler than the one provided before~\cite{tangen2007} and we have introduced a new constant $U_0$ in our formulation of the JRW transformation.
This new constant is missing in the original four-dimensional formulation~\cite{jrw1969} but crucial for providing a new integration constant.
As a complement, we have also presented alternative derivations of the higher-dimensional Buchdahl and JRW transformations based on the reduced action for static spacetimes in the form of a nonlinear sigma model.
Lastly, adopting the $n(\ge 4)$-dimensional JRW transformation to the generalized JNW solution with an Einstein base manifold, we have shown that the resulting charged solution is the type-I solution for $k=1,-1$ and the type-VI${}_0$ solution for $k=0$ in the complete classification performed in~\cite{mm2016}.
As other demonstrations, we have constructed two new electrically charged solutions.
One is an $n$-dimensional multi-center solution with a ghost scalar field and the other is a cylindrically symmetric hairy solution in four dimensions.

Combining the $n$-dimensional Buchdahl transformation and a conformal transformation to a solution in the present system without a Maxwell field, one can construct various static solutions in the system with a nonminimally coupled scalar field.
This is also the case in $f(R)$ gravity because such a system is conformally transformed into general relativity with a minimally coupled scalar field~\cite{maeda1989}.
Similarly, one can combine the JRW transformation and a conformal transformation in four dimensions to construct a charged solution in other systems, but this is not allowed in higher dimensions because then the Maxwell field is not invariant under conformal transformations.
An interesting example in this context is the so-called BBMB spherically symmetric solution in four dimensions~\cite{Bocharova:1970skc,Bekenstein:1974sf} with a conformally coupled scalar field.
This solution is obtained from the JNW solution by a conformal transformation and represents an extremal black hole with a scalar hair.
Interestingly enough, its higher-dimensional counterpart represents not a black hole but a naked singularity~\cite{klimcik1993}.

This kind of properties of spacetime dimensionality is also observed in the $n$-dimensional Majumdar-Papapetrou solution which represents multi-black holes only in four and five dimensions~\cite{cr2007}.
These two examples surely suggest that there is a close relation between spacetime configurations and dimensionality of spacetime.
In the present paper, we have constructed an electrically charged multi-center solution with a massless ghost scalar field.
In the neutral case, this solution can represent multi-wormholes in four dimensions~\cite{eks2016,gv2017}.
However, since the differentiability of spacetime could sharply depend on the number of dimensions, a careful and intensive analysis is required in order to provide a correct physical interpretation of the solution in higher dimensions or with electric charge. We leave this problem for future investigations.

\subsection*{Acknowledgements}
The authors thank Stoytcho S.~Yazadjiev for bringing our attention to the nonlinear sigma model approach.
The authors also thank Yen-Kheng Lim, Shinya Tomizawa, and Jorge Zanelli for helpful comments.
C.~M.~thanks Hokkai-Gakuen University for a kind hospitality, where this work was completed.
This work has been partially funded by the Fondecyt
grants 1161311 and 1180368. The Centro de Estudios Cient\'{\i}ficos (CECs) is funded by the Chilean Government through the Centers of Excellence Base Financing Program of Conicyt.

\appendix

\section{Static decompositions}
\label{App:derivation}
In this appendix, we derive the field equations~(\ref{ein-eq1})--(\ref{matter-eq}).
Let us consider $n(\ge 4)$-dimensional static spacetimes ${\ma M}^n \approx \mathbb{R} \times M^{n-1}$ with the following general metric 
\begin{eqnarray}
g_{\mu\nu}\D x^\mu \D x^\nu=-\Omega(y)^{-2}\D t^2+g_{ij}(y)\D y^i\D y^j, \label{eq:structure}
\end{eqnarray}
where the indices $i$ and $j$ run from $1$ to $n-1$ and $g_{ij}$ and $\Omega$ are an arbitrary Euclidean metric and a scalar function on $M^{n-1}$, respectively. 
We introduce the covariant derivative $M^{n-1}$ such that $D_kg_{ij}=0$.
Then, non-zero components of the Christoffel symbol in the spacetime (\ref{eq:structure}) are 
\begin{eqnarray}
{}^{(n)}{\Gamma}{}^t_{ti}=-\Omega^{-1}D_i\Omega, \qquad {}^{(n)}{\Gamma}{}^i_{tt}=-\Omega^{-3}D^i\Omega, \qquad {}^{(n)}{\Gamma}{}^i_{jk}={}^{(n-1)}{\Gamma}{}^i_{jk}, \label{Gamma-static}
\end{eqnarray}  
where the superscripts $(n)$ and $(n-1)$ imply geometrical quantities on ${\ma M}^n$ and $M^{n-1}$, respectively.
By Eqs.~(\ref{Gamma-static}), non-zero components of the curvature tensors are calculated to give
\begin{align}
{}^{(n)}{\cal R}_{ijkl}=&{}^{(n-1)}{\cal R}_{ijkl},\\
{}^{(n)}{\cal R}_{titj}=&2\Omega^{-4}(D_i\Omega)(D_j\Omega)-\Omega^{-3}D_iD_j\Omega,\\
{}^{(n)}{\cal R}{}_{tt}=&2\Omega^{-4}(D\Omega)^2-\Omega^{-3}D^2\Omega,\\
{}^{(n)}{\cal R}{}_{ij}=&{}^{(n-1)}{\cal R}{}_{ij}-2\Omega^{-2}(D_i\Omega)(D_j\Omega)+\Omega^{-1}D_iD_j\Omega,\\
{}^{(n)}{\cal R}=&{}^{(n-1)}{\cal R}-4\Omega^{-2}(D\Omega)^2+2\Omega^{-1}D^2\Omega,
\end{align} 
where $(D\Omega)^2:=g^{ij}(D_i\Omega)(D_j\Omega)$ and $D^2\Omega:=g^{ij}D_iD_j\Omega$.
Using the above decompositions, we obtain the following non-zero components of the Einstein tensor:
\begin{align}
{}^{(n)}{G}{}_{tt}=&\frac12\Omega^{-2}({}^{(n-1)}{\cal R}),\label{Ein-decomp1}\\
{}^{(n)}{G}{}_{ij}=&{}^{(n-1)}{G}{}_{ij}-2\Omega^{-2}(D_i\Omega)(D_j\Omega)+\Omega^{-1}D_iD_j\Omega+g_{ij}\left\{2\Omega^{-2}(D\Omega)^2-\Omega^{-1}D^2\Omega\right\}.\label{Ein-decomp2}
\end{align}

Now we consider a conformally related $(n-1)$-dimensional Euclidean metric ${\bar g}_{ij}$ such that ${g}_{ij}(y)=\Omega(y)^{2p}{\bar g}_{ij}(y)$, where $p$ is a constant.
In terms of ${\bar g}_{ij}$, ${}^{(n-1)}{\cal R}_{ij}$, ${}^{(n-1)}{\cal R}$, and ${}^{(n-1)}{G}_{ij}$ are written as
\begin{align}
{}^{(n-1)}{\cal R}_{ij}=&{}^{(n-1)}{\bar {\cal R}}_{ij}-p(n-3){\bar D}_i {\bar D}_j\ln\Omega-p{\bar g}_{ij}{\bar D}^2 \ln\Omega \nonumber  \\
&+(n-3)p^2({\bar D}_i\ln\Omega)({\bar D}_j\ln\Omega)-p^2(n-3){\bar g}_{ij}({\bar D}\ln\Omega)^2,\\
{}^{(n-1)}{\cal R}=&\Omega^{-2p}\left\{{}^{(n-1)}{\bar {\cal R}}-2p(n-2){\bar D}^2\ln\Omega-p^2(n-2)(n-3)({\bar D}\ln\Omega)^2\right\},\label{scalarRbarR}\\
{}^{(n-1)}{G}_{ij}=&{}^{(n-1)}{\bar G}_{ij}-p(n-3){\bar D}_i {\bar D}_j\ln\Omega+p^2(n-3)({\bar D}_i\ln\Omega)({\bar D}_j\ln\Omega) \nonumber \\
&+p(n-3){\bar g}_{ij}{\bar D}^2\ln\Omega+\frac12p^2(n-3)(n-4){\bar g}_{ij}({\bar D}\ln\Omega)^2,
\end{align}
where geometrical quantities with bars are the ones constructed from ${\bar g}_{ij}$~\cite{wald}.
The second derivatives of $\Omega$ on $M^{n-1}$ are written as
\begin{align}
D_iD_j\Omega=&{\bar D}_i{\bar D}_j\Omega-p\Omega^{-1}\biggl\{2({\bar D}_i \Omega)({\bar D}_j\Omega)-{\bar g}_{ij}({\bar D} \Omega)^2\biggl\},\\
D^2\Omega=&\Omega^{-2p}{\bar D}^2\Omega+p(n-3)\Omega^{-2p-1}({\bar D} \Omega)^2.
\end{align}
By the above relations with $p=1/(n-3)$, the Einstein tensor~(\ref{Ein-decomp1}) and (\ref{Ein-decomp2}) reduce to
\begin{align}
{}^{(n)}{G}{}_{tt}=&\frac12\Omega^{-2(n-2)/(n-3)}\left\{{}^{(n-1)}{\bar {\cal R}}-\frac{2(n-2)}{n-3}{\bar D}^2\ln\Omega-\frac{n-2}{n-3}({\bar D}\ln\Omega)^2\right\},\label{Ein-decomp3} \\
{}^{(n)}{G}{}_{ij}=&{}^{(n-1)}{\bar G}_{ij}-\frac{n-2}{n-3}({\bar D}_i\ln\Omega)({\bar D}_j\ln\Omega) +\frac{n-2}{2(n-3)}{\bar g}_{ij}({\bar D}\ln\Omega)^2.\label{Ein-decomp4}
\end{align}

In order to write down the field equations in terms of ${\bar g}_{ij}$, we assume $\phi=\phi(y)$ and $A_\mu=A_t(y)\delta^t_\mu$.
Under these assumptions, the massless Klein-Gordon equation and the Maxwell equation given by Eq.~(\ref{em-kg}) are written as
\begin{align}
{\bar D}^2\phi=0,\qquad \partial_i(\sqrt{\det{\bar g}}\Omega^2{\bar g}^{ij}F_{tj})=0.
\end{align}
Also, non-zero components of the energy-momentum tensors (\ref{Tab-Max}) and (\ref{Tab-scalar}) are
\begin{align}
T^{(\rm em)}_{tt}=&\frac12\Omega^{-2/(n-3)}{\bar g}^{kl}F_{tk}F_{tl}, \label{Tab-Max-decomp1}\\
T^{(\rm em)}_{ij}=&-\Omega^2\biggl(F_{it}F_{jt}-\frac 12 {\bar g}_{ij}{\bar g}^{kl}F_{tk}F_{tl}\biggl)\label{Tab-Max-decomp2}
\end{align}
and 
\begin{align}
T^{(\phi)}_{tt} =&\frac12\Omega^{-2(n-2)/(n-3)}({\bar D}\phi)^2,\label{Tab-scalar-decomp1}\\
T^{(\phi)}_{ij} =&({\bar D}_i \phi)({\bar D}_j \phi)-\frac12{\bar g}_{ij}({\bar D}\phi)^2.\label{Tab-scalar-decomp2}
\end{align}

By Eqs.~(\ref{Ein-decomp3}), (\ref{Ein-decomp4}), (\ref{Tab-Max-decomp1})--(\ref{Tab-scalar-decomp2}), the Einstein equations (\ref{EFE}) are finally written as
\begin{align}
&{}^{(n-1)}{\bar {\cal R}}-\frac{2(n-2)}{n-3}{\bar D}^2\ln\Omega-\frac{n-2}{n-3}({\bar D}\ln\Omega)^2=\kappa_n\biggl\{({\bar D}\phi)^2+\Omega^2{\bar g}^{kl}F_{tk}F_{tl}\biggl\},\\
&{}^{(n-1)}{\bar G}_{ij}-\frac{n-2}{n-3}({\bar D}_i\ln\Omega)({\bar D}_j\ln\Omega) +\frac{n-2}{2(n-3)}{\bar g}_{ij}({\bar D}\ln\Omega)^2 \nonumber \\
&~~~~~~~=\kappa_n\biggl\{({\bar D}_i \phi)({\bar D}_j \phi)-\frac12{\bar g}_{ij}({\bar D}\phi)^2-\Omega^2\biggl(F_{it}F_{jt}-\frac 12 {\bar g}_{ij}{\bar g}^{kl}F_{tk}F_{tl}\biggl)\biggl\},
\end{align}
which reduce to
\begin{align}
&{}^{(n-1)}{\bar {\cal R}}_{ij}-\frac{n-2}{n-3}({\bar D}_i\ln\Omega)({\bar D}_j\ln\Omega) =\kappa_n\biggl\{({\bar D}_i \phi)({\bar D}_j \phi)-\Omega^2F_{it}F_{jt}\biggl\},\\
&-\frac{n-2}{n-3}{\bar D}^2\ln\Omega=\kappa_n\Omega^2{\bar g}^{kl}F_{tk}F_{tl}.
\end{align}

\end{document}